\documentclass[12pt]{article}
\usepackage{fleqn,espcrc1,graphicx}

\title{Review of Nuclear Reactions at the AGS}
\author{C.A. Ogilvie\address{Physics and Astronomy Department,
Iowa State University, Ames, IA 50011}}

\begin{document}

\maketitle

\begin{abstract}
Results from p+A and A+A collisions from the beam energies 2-18 AGeV/c
are reviewed with emphasis on the properties of dense hadronic matter,
and its implications for claims that a new state of matter
has been formed at the SPS. 

\end{abstract}

\section{Introduction}

The central goal of  AGS heavy-ion program is the study of 
hadronic matter at several times normal nuclear density:
how this matter is formed, and the characterization of 
its properties. The physics of dense matter is compelling
in its own right. In addition we need
to understand  how the properties of dense matter 
affects QGP signatures, to help assess whether the QGP 
has been formed in heavy-ion collisions\cite{qgpclaim}.
My goals for this talk are to review what we have learned about dense 
hadronic matter, 
to assess the open questions, and to sketch the
possibilities for a future research program. 

\section{Mean-fields in Dense Matter}

\begin{figure}[htb]
	\centering
	\includegraphics[width=3.0in]{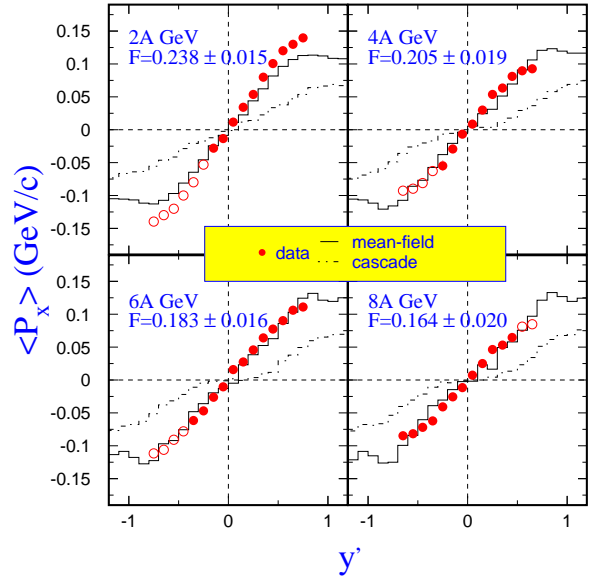}
	\caption{The measured $<p^x>$  for protons
	from Au+Au reactions as a function of beam energy from
	E895\protect{\cite{e895flow}}.
	The dashed line is a cascade calculation and the
	solid line includes a repulsive mean-field.}
	\label{fig:e895flow}
\end{figure}

The mean free-path of a typical hadron 
in dense matter is $\lambda=1/\rho\sigma$, hence at four times
normal nuclear density, $\lambda\sim$0.5 fm. This mean free-path is 
so short that successive collisions are unlikely to be independent
processes. Attempts have been made to model n-body 
collisions\cite{nbody} as well as off mass-shell propagation between
collisions\cite{offmass}. A simpler ansatz may be to 
incorporate the average effects of multiple
soft interactions into an effective mean-field. 
This extends a long tradition in low-energy nuclear 
physics to higher densities.
Within this mean-field,
successive elastic and inelastic
two-body collisions between hadrons take place, forming resonances
and strings that subsequently decay.

One test of this mean-field ansatz is the measurement of the 
average transverse momentum in the reaction plane ($<p^x>$).
Protons measured by
E895\cite{e895flow}
flow within the reaction plane , i.e. 
have opposite sign $<p^x>$ either side of 
mid-rapidity (Figure~\ref{fig:e895flow}).
Qualitatively
the flow of protons is driven by pressure
gradients in the reaction zone.  Quantitatively
the magnitude of the flow is not reproduced by calculations that
consider only a cascade of two-body collisions. 
Much better agreement is obtained by including
the additional deflection driven by
gradients in the repulsive nuclear mean-field.
\begin{figure}[htb]
	\begin{minipage}[l]{0.5\textwidth}
		\centering
		\includegraphics[width=3.2in]{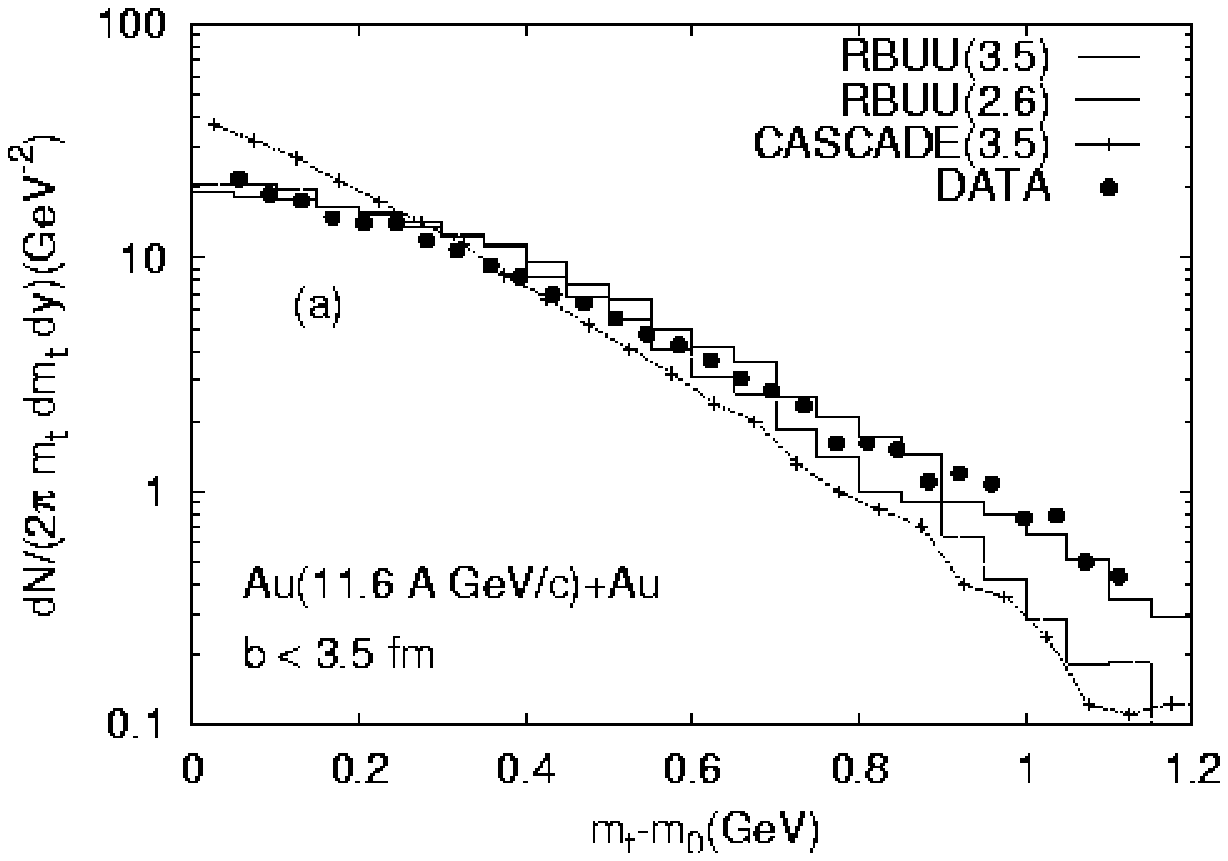}

	\end{minipage}
\
	\begin{minipage}[r]{0.5\textwidth}
		\centering
		\includegraphics[width=3.0in]{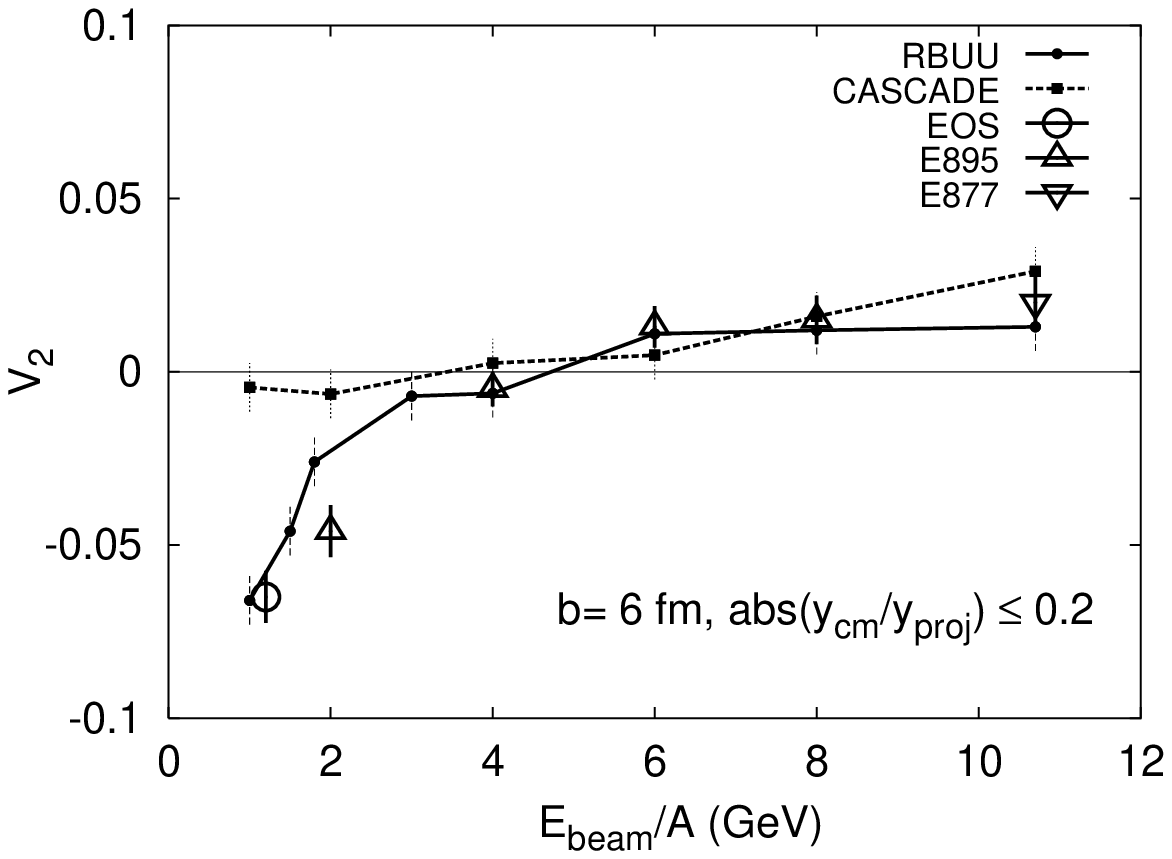}
	
		\end{minipage}
	\caption{The left panel shows
	the invariant yield of protons at mid-rapidity
	from Au+Au at 11.6AGeV/c\protect{\cite{e866proton}}, 
	while the right panel shows the
 	measured elliptic flow v$_2$ from Au+Au(open symbols).
	Both observables are well
	reproduced by  transport calculation (RBUU 3.5) 
	that include a momentum-dependent
	repulsive mean-field\protect{\cite{Sahu00}}.} 
	\label{fig:meanField}
\end{figure}		

The proposed mean-field has a complicated dependence on density and
momentum\cite{cassingMF}. At low momenta the field is moderately 
repulsive at high-densities, then as the  momentum of the particle
increases, the field progressively becomes more repulsive.
Such a momentum dependent 
mean-field was 
found to be necessary\cite{Sahu00}
to reproduce the measured p$_t$ distributions of proton from Au+Au
reactions at 11.7 AGeV/c (Figure~\ref{fig:meanField})\cite{e866proton}.

A sensitive check on the
applicability of a mean-field at high densities is elliptic flow: the 
back-to-back azimuthal emission of particles.
At low beam energies  elliptic flow is oriented 90$^{\circ}$ with
respect to the reaction plane, where the matter is
"squeezed" out from the top and bottom of the reaction zone.
For squeezeout, the second-order moment v$_2$ 
has negative values. 
As the 
beam energy increases, the spectator matter moves rapidly away 
from the collision, which decreases
the spectators' effectiveness at blocking mid-rapidity emission. 
The beam energy dependence of elliptic flow is therefore
sensitive to pressure gradients and the reaction
time-scale\cite{SorgeElliptic}. 
The elliptic flow data from E895\cite{E895Elliptic} and E877\cite{E877Elliptic}
(figure \ref{fig:meanField}) change from being directed 90$^{\circ}$ 
to the reaction plane (negative moment v$_2$)
to elliptic-flow within the reaction plane (positive moment v$_2$).
This evolution, as well as the location
of no elliptic flow near 4 AGeV,
is quantitatively reproduced by a transport
model that includes the repulsion
due to a momentum dependent mean-field\cite{Sahu00}.

\begin{figure}[htb]
	\centering
	\includegraphics[width=4.0in]{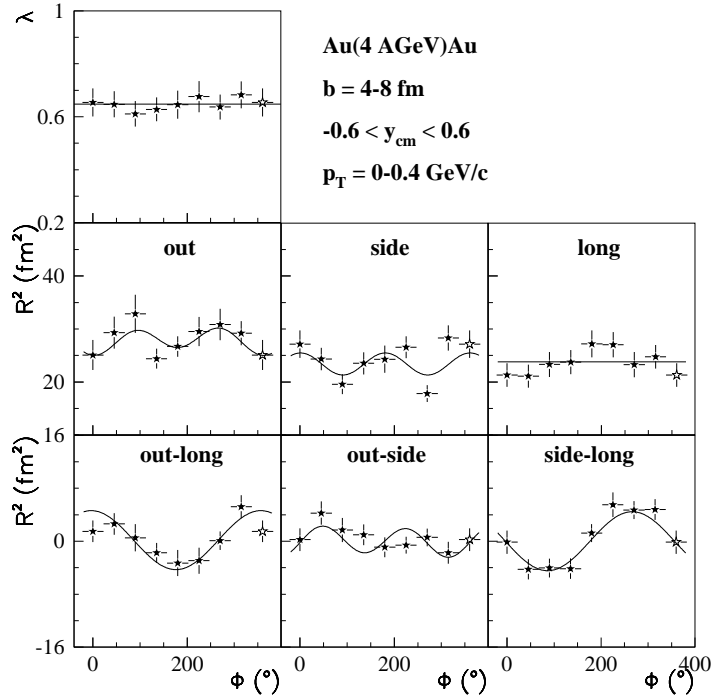}
	\caption{The extracted HBT parameters as a function of the
	azimuthal angle between the pair and the reaction plane
	for Au+Au at 4 AGeV\protect{\cite{e895hbt}}. 
	The lines are a global fit.}
	\label{fig:hbtAzimuth}	
\end{figure}

A future test of the mean-field dynamics will be whether 
transport models can reproduce the measured azimuthal dependence of
pion correlations. Shown in Figure \ref{fig:hbtAzimuth} are the
extracted HBT parameters measured as a function of azimuthal angle from
Au+Au reactions at 4 AGeV\cite{e895hbt}.
As an indication of
the richness of the information in this data, consider the 
case when $\phi$=0, i.e. when the transverse momentum
of the pair is parallel to the reaction plane. At $\phi$=0
the data indicate that 
$R_{out} < R_{side}$ consistent with the almond shape of the
initial collision zone. It is however remarkable that this shape 
information is retained throughout the dynamics  of the reaction to
freezeout. 
\clearpage

\begin{figure}[htb]
	
	\centering 
	\includegraphics[width=3.4in]{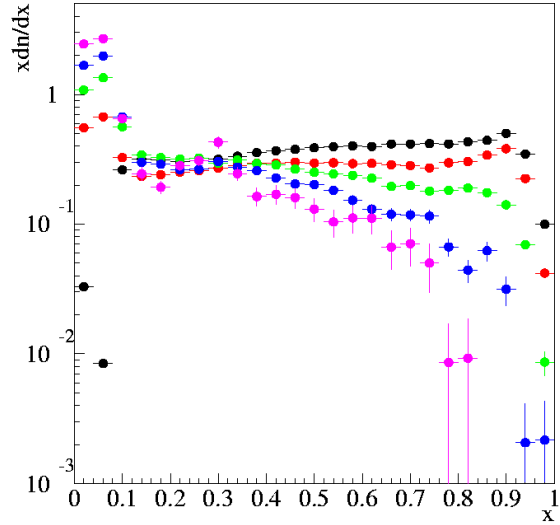}	
	\caption{The preliminary x-distribution of protons from p+Be
	reactions at 12 GeV/c\protect{\cite{coleqm2001}}. 
	Reactions with number of collisions
	$\nu\sim 1$ have the largest yield at high x. 
	The x-distributions steepen as 
	$\nu$ increases to $\nu\sim 2.3$.}
	\label{fig:xproton}
\end{figure}

The major caveat in these tests of the high-density 
mean-field is whether the multiple elastic
and inelastic collisions
are under control in transport models. Recent data on proton
distributions in p+A reactions 
provide insight into these scattering processes, 
and therefore act as a critical benchmark. 
The x-distributions (x$=p_z/p_{beam}$)
of protons from p+Be reactions at 18 GeV/c\cite{coleqm2001} are shown in 
figure \ref{fig:xproton}. In  reactions where the mean number of
collisions ($\nu$) is close
to 1, the measured protons are broadly
spread over a wide range of x-values. 
The measured proton x-distribution shifts backwards as the number 
of collisions increases, consistent with more momentum exchange in
multiple collisions. This evolution provides a quantitative
test of a transport model's description of 
rescattering after the first collision. 

\begin{figure}
	\centering
	\includegraphics[width=3.0in]{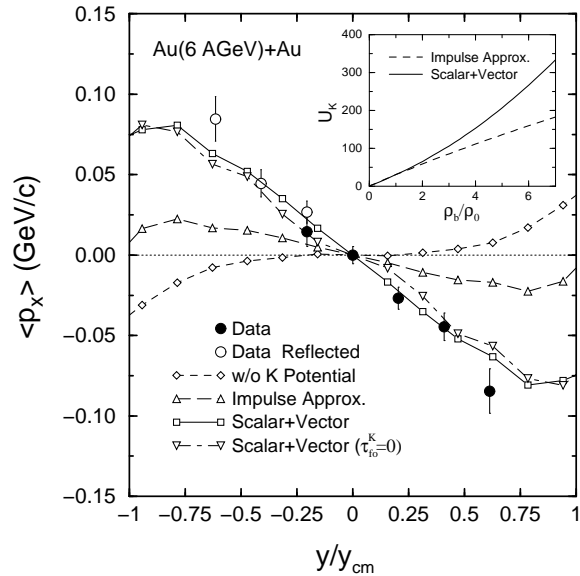}
	\caption{The measured K$^0_s$ flow in solid symbols from Au+Au
	reactions at 6AGeV\protect{\cite{K0barFlow}} 
	is compared with transport models. The
	kaon potential is shown as an insert.}
	\label{fig:k0flow}
	
\end{figure}

Another open question is whether mesons experience
a mean-field in dense matter.
Data on kaonic atoms\cite{kaonAtom} suggest that K$^-$
have attractive resonant interactions with baryons, 
whereas for K$^+$ and K$^0$
the interaction with baryons is predicted to be repulsive.
The interactions
kaons experience in dense matter can be incorporated into a 
mean-field or expressed as a change in the dispersion equation of a kaon 
propagating in dense matter\cite{SchaffnerKaonMass}. 
The  energy at zero momentum is
the kaon's effective mass, which for
K$^-$ is predicted to decrease 
at large densities\cite{SchaffnerKaonMass}. 
The cleanest 
observation of the repulsive mean-field for K$^0$ is the 
directed transverse momentum for K$^0_s$\cite{kaonPotential,K0barFlow}
which is opposite that for the
baryons (figure \ref{fig:k0flow}). 
The data are well reproduced by a transport model that include
a vector and scalar
mean-field that repels the K$^0_s$ away from the baryons.

\section{Rescattering in Dense Matter}

The dense environment in a heavy-ion collision
changes the yield of produced particles as compared to
p+p reactions.
As an example, the centrality dependence of particle production in Si+Au and
Au+Au reactions, shows a strong increase in the
kaon yield {\it per participant nucleon}\cite{e866char} 
from peripheral
to central reactions.
This is consistent with rescattering
in central reactions being more effective at producing kaons. 
The yield of pions per participating nucleon changes more slowly.

\begin{figure}[htb]
	
	\begin{minipage}[l]{0.5\textwidth}
	  \centering 
		\includegraphics[width=3.2in]{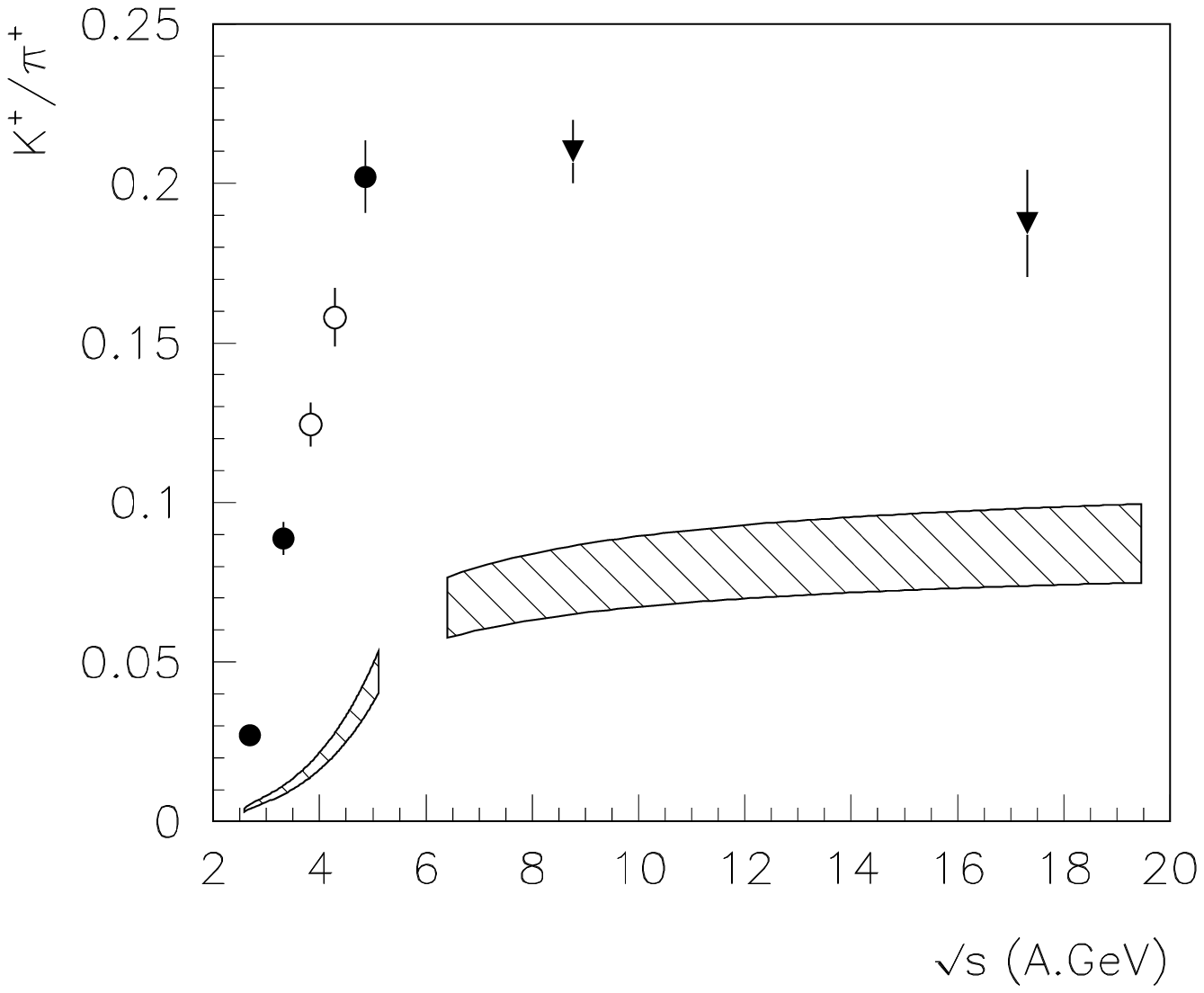}	
	\end{minipage}%
	\begin{minipage}[r]{0.5\textwidth}
		\centering
		\includegraphics[width=3.2in]{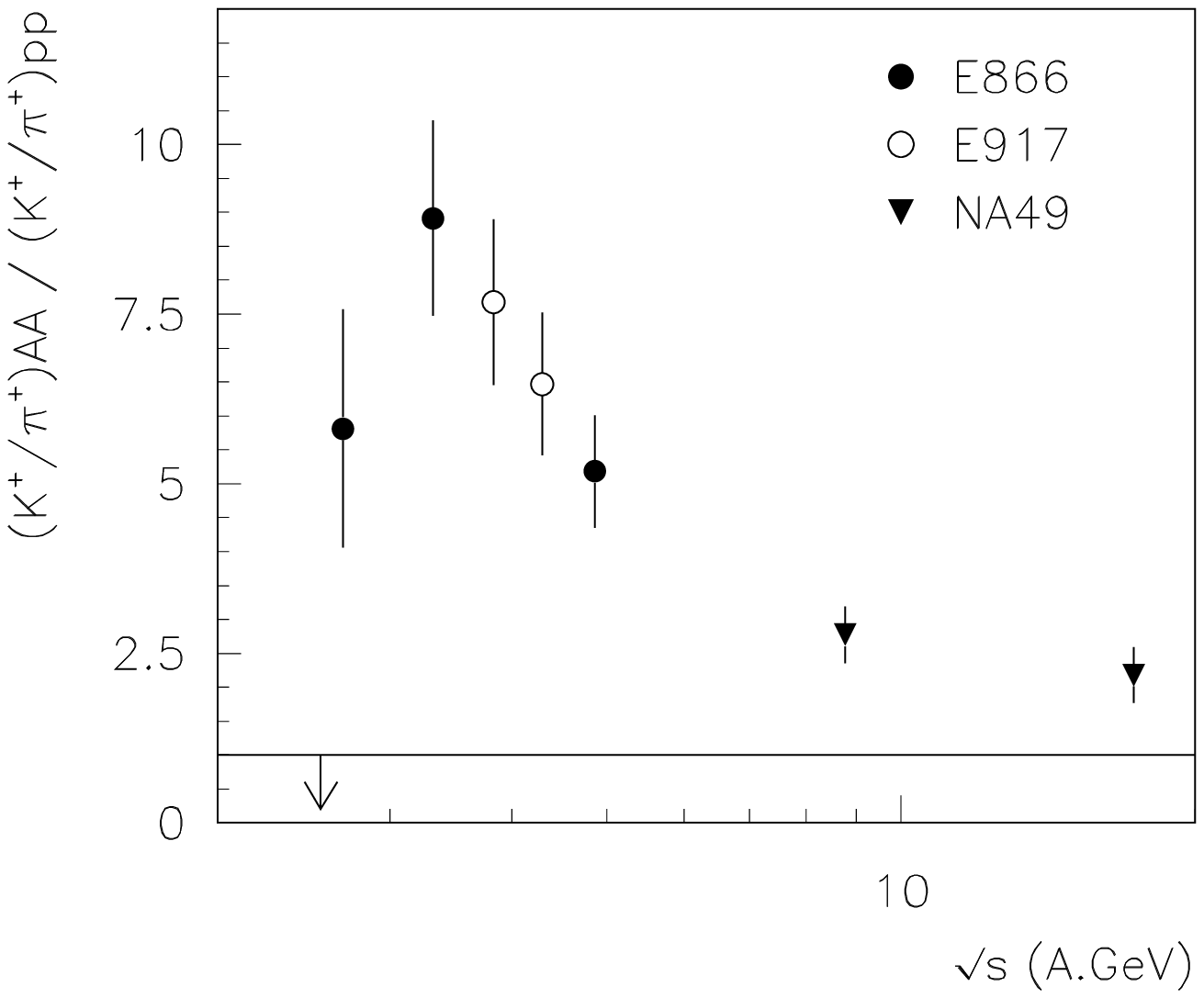}
	
	\end{minipage}%
	\caption{Left panel shows the measured $K/\pi$ ratio from central
		collisions as a function of beam 
		energy\protect{\cite{kpiEbeam}}. The hashed region
		represents the $K/\pi$ ratio from p+p reactions.
		Right panel shows the
		enhancement of $K/\pi$ ratio (Au+Au ratio/p+p ratio).}
		\label{fig:kpiEbeam}
	
\end{figure}

These increases in particle production depend complexly on
the number of rescattering collisions during a heavy-ion reaction 
and on the energetics and type of each rescattering.
The balance between these factors
changes as the beam energy is lowered. 
For example, the measured $K/\pi$ ratio\cite{kpiEbeam} 
(Figure \ref{fig:kpiEbeam})
decreases as the beam
energy is reduced. However the decrease is not as rapid
as the  decrease in $K/\pi$ ratio from p+p reactions. 
This is most easily seen in the $K/\pi$ enhancement
shown in figure \ref{fig:kpiEbeam}. The enhancement is largest at the lower
beam energies which indicates that rescattering becomes relatively 
more important as the beam energy decreases. At low
energies, secondary collisions  are
close to the kaon production threshold, so the increased
enhancement implies a large increase in the
number of secondary collisions.

It is worth noting that
moving in the other direction and
increasing the beam energy, the $K/\pi$ enhancement smoothly decreases 
from AGS energies to the
SPS at 40 and 160 AGeV/c. This is consistent with a smooth evolution of the 
reaction mechanism\cite{jcdunlop00}. 
Furthermore the maximum in the $K/\pi$ ratio in A+A reactions
can be explained as a mathematical artifact
caused by the convolution of a rising $K/\pi$ ratio from p+p reactions and 
a falling enhancement as the beam energy increases\cite{jcdunlop00}.

\begin{figure}[htb]
	
	\begin{minipage}[l]{0.5\textwidth}
		  \includegraphics[width=3.1in]{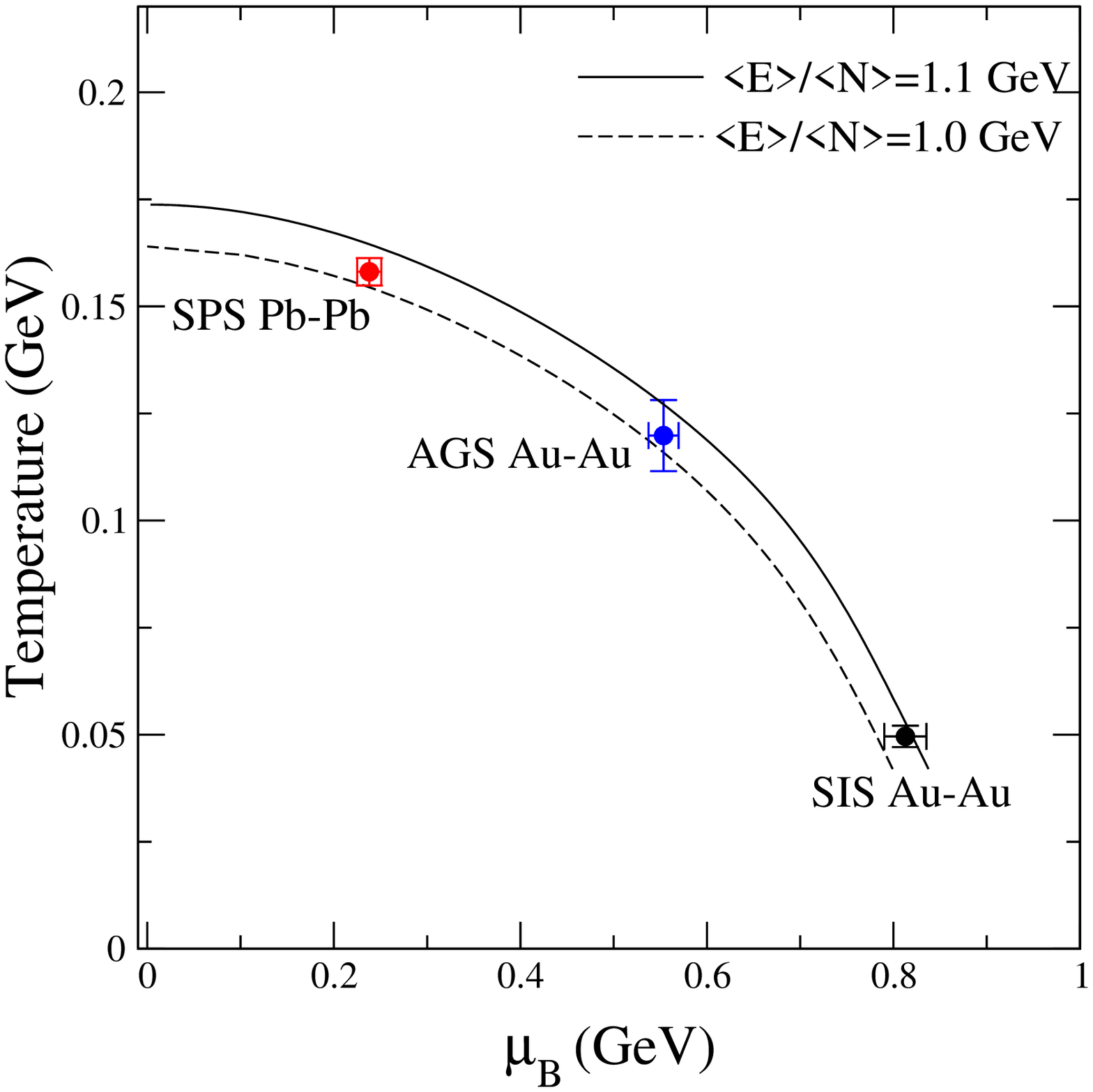}	
	\end{minipage}%
	\begin{minipage}[r]{0.5\textwidth}
		 \includegraphics[width=3.1in]{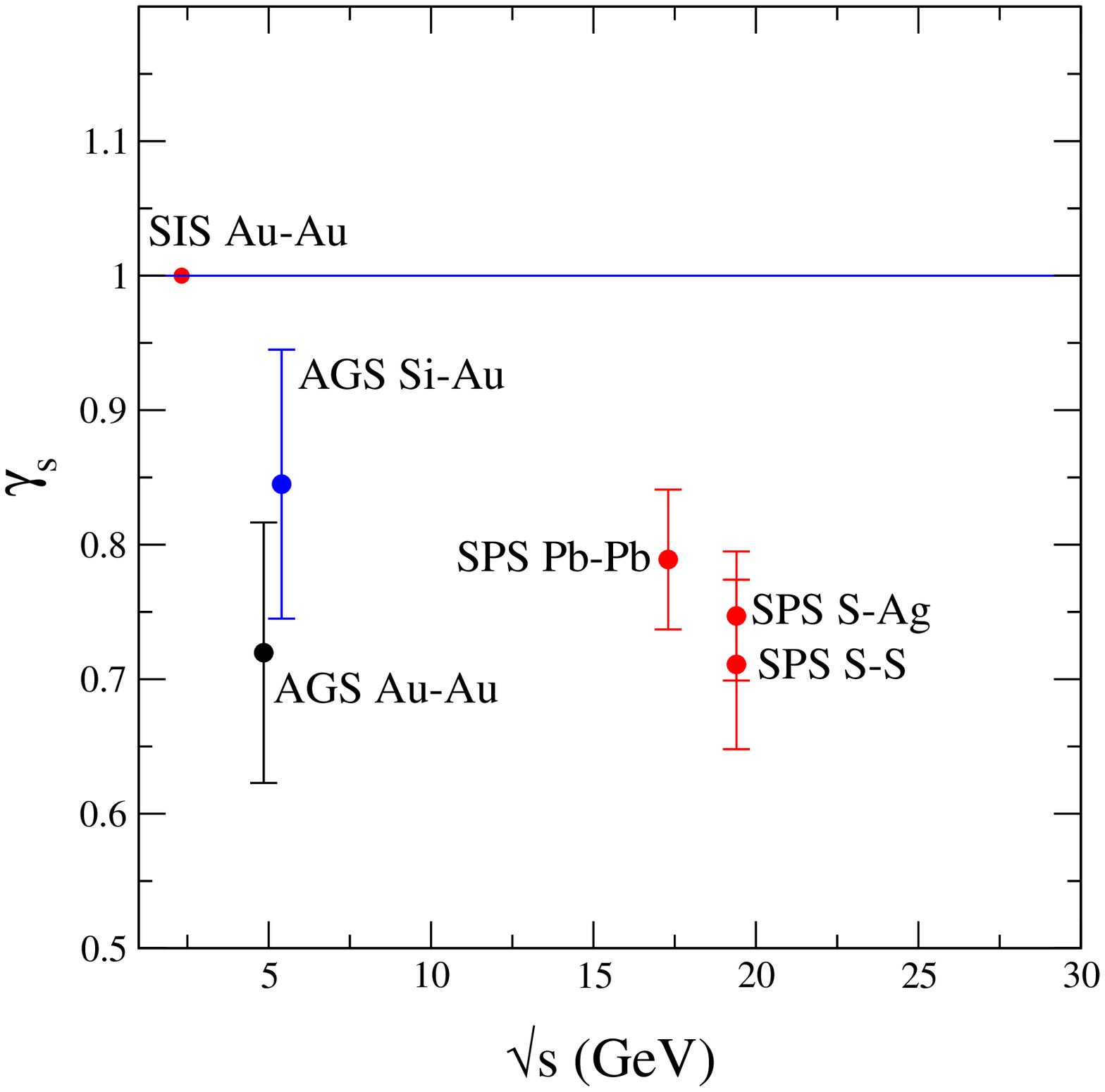}
	\end{minipage}%

	\caption{Thermal parameters fit to
	data from heavy-ion reactions from 1-160 AGeV.}
	\label{fig:chemTmus}

\end{figure}

A sufficiently large rescattering rate  
 will drive the dense matter towards
thermal equilibrium. Thermal models have been very successful
in reproducing particle ratios from A+A collisions\cite{becattini00}.
The fit parameters of these models, temperature and baryon
chemical potential, map out a contour of constant energy 
density\cite{becattini00} (figure \ref{fig:chemTmus}).
The third parameter in the model is the strangeness saturation 
factor ($\gamma_s$) which measures how close 
the system is to complete chemical 
equilibrium. As can be seen in Figure \ref{fig:chemTmus},
$\gamma_s$ is approximately 0.7 in heavy-ion reactions 
at both AGS\cite{noMultipleStrange} and SPS energies. This is
larger than the $\gamma_s \sim 0.2$ extracted from p+p 
reactions\cite{becattini00}, 
indicating that there are strong processes in heavy-ion
collisions that drive the system
towards full chemical equilibrium.

\begin{figure}
	
	\begin{minipage}[l]{0.5\textwidth}
	\centering
	   \includegraphics[width=3.2in]{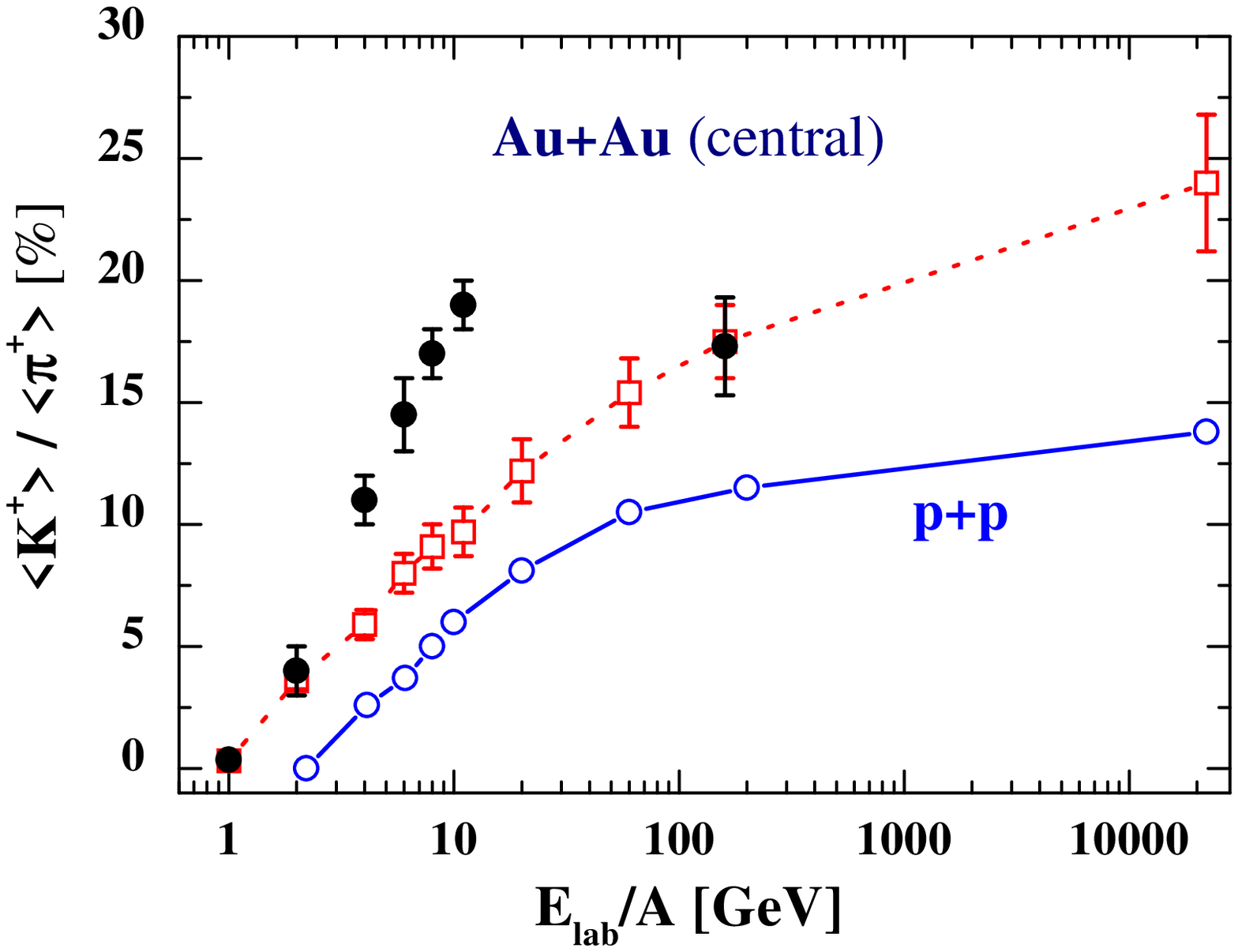}	
	\end{minipage}%
	\begin{minipage}[r]{0.5\textwidth}
		\centering
		\includegraphics[width=3.0in]{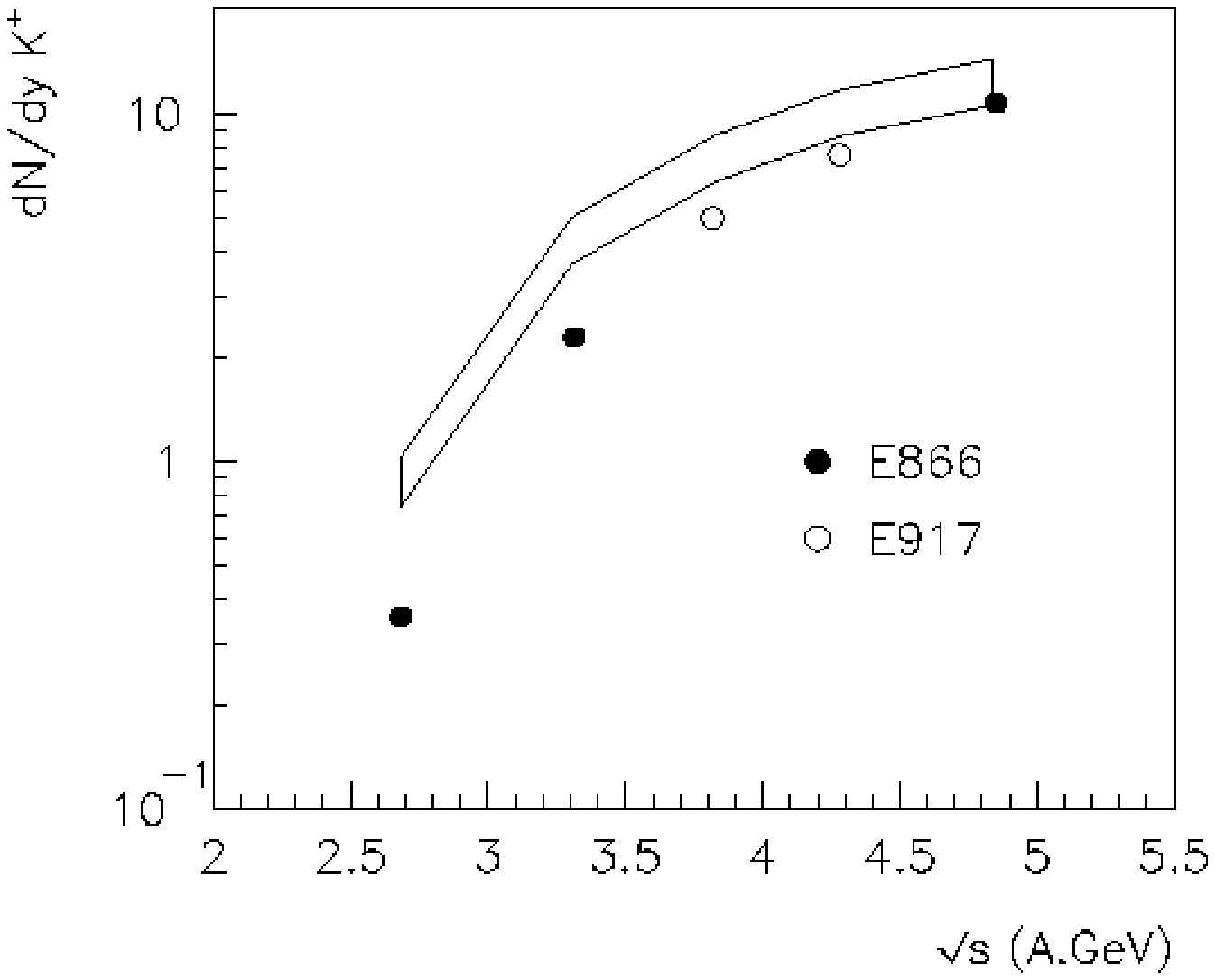}
		
	\end{minipage}%
	\caption{In the left panel the measured K/$\pi$
	ratio (solid symbols) is compared with a calculation from HSD transport
	model (open squares), 
	and on the right is a comparison between the mid-rapidity
	K$^+$ yield and a RQMD calculation (open box).}
	\label{fig:kaonModel}
\end{figure}

At SPS energies it has been suggested that 
the 
increase in strangeness production occurs in a new state of 
matter, possibly the QGP\cite{qgpclaim}. At AGS energies it is widely
considered that the rescattering of hadrons drives the system towards
strangeness saturation. 
It is unsatisfactory to have two separate
explanations for very similar experimental
results and we need to be confident that
hadronic rescattering cannot explain the strangeness production
at the SPS.
The difficulty is illustrated by  
two 
comparisons between hadronic cascade transport calculations
and measured strangeness production that are shown in 
Figure \ref{fig:kaonModel}. The Hadron String Dynamics (HSD) 
model\cite{cassingMF} underpredicts the AGS K$^+/\pi^+$ data,
while the RQMD cascade model reproduces the K$^+$ yield at the 
highest AGS energy but increasingly overpredicts 
the data as the beam energy is reduced.
Based on this trend, it is not
surprising that RQMD underpredicts the strangeness yield at SPS 
energies\cite{Odyniec,noMultipleStrange}. Given the broad and systematic
failure of these models to describe strangeness production at
any energy, it does not seem prudent to use their
underprediction of strangeness to support
the argument that the QGP was formed at SPS energies.
We need to understand the failure of these models
at 1-10 AGeV/c to help interpret the failure of
these models at 160 AGeV/c.

\begin{figure}[htb]
	
	\begin{minipage}[c]{0.5\textwidth}
	   \includegraphics[width=3.0in]{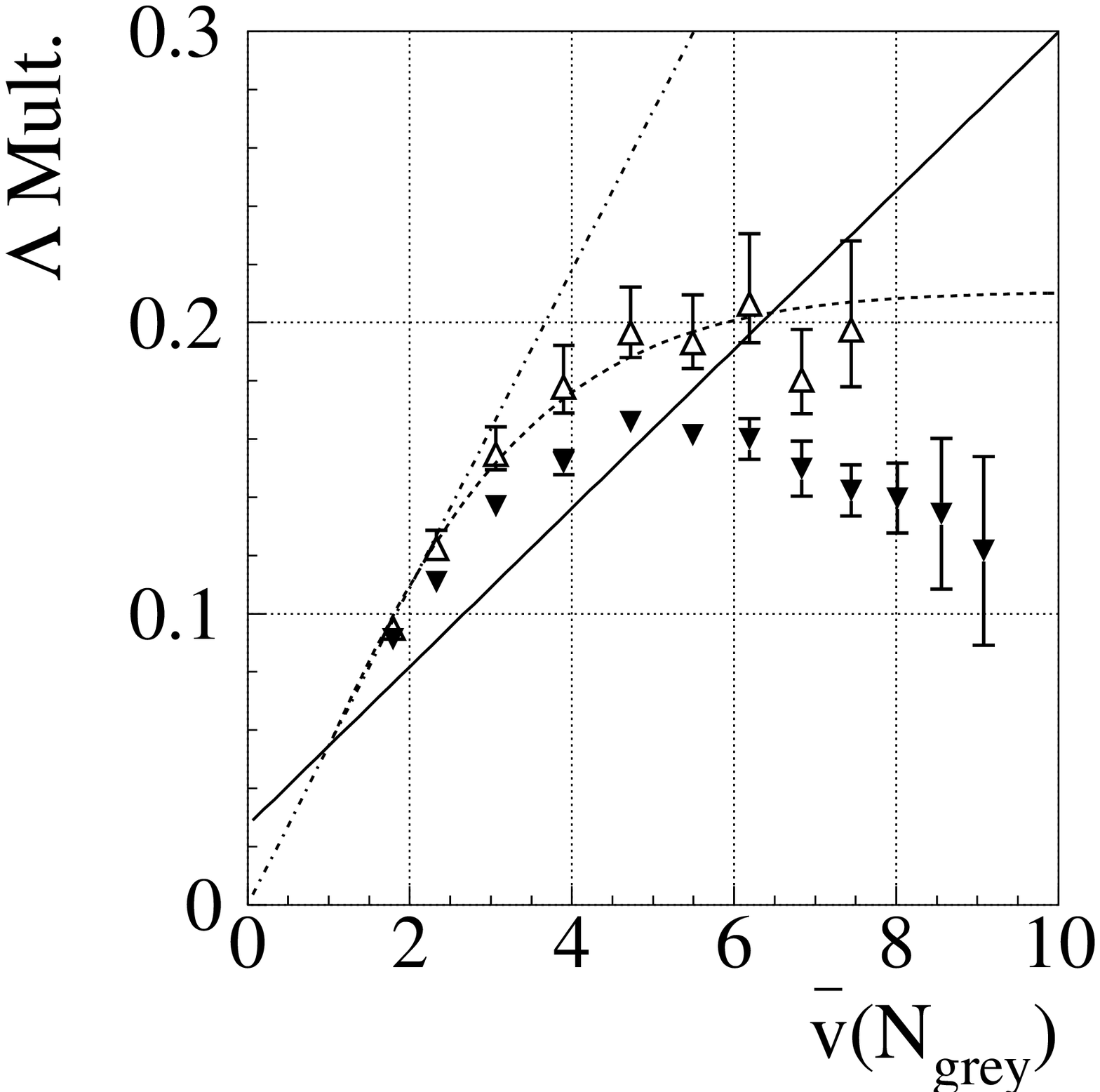}	
	\end{minipage}%
	\begin{minipage}[c]{0.5\textwidth}
	   \includegraphics[width=3.0in]{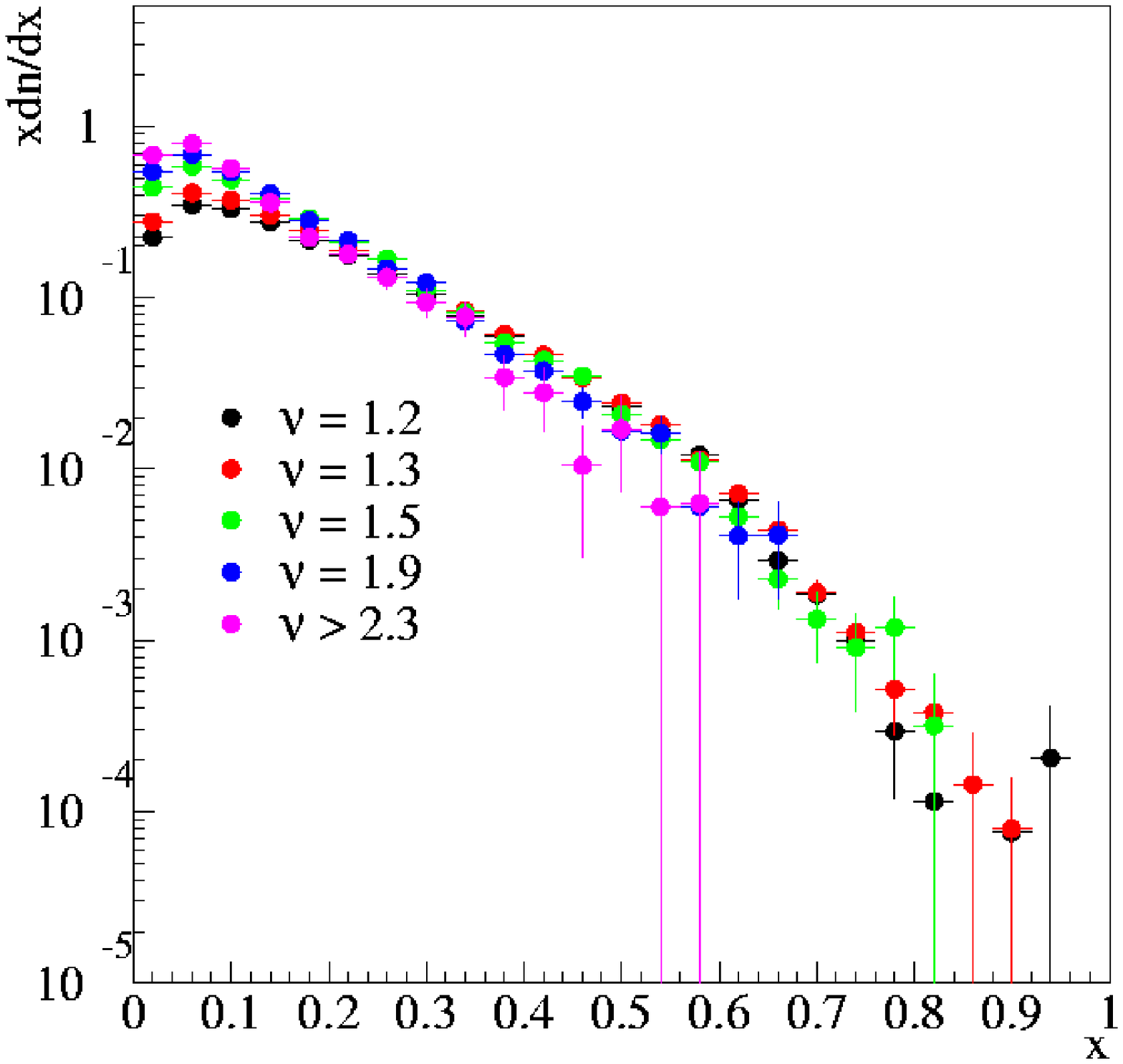}
	\end{minipage}%
	\caption{Left panel shows the total yield (open symbols)
	 of $\Lambda$ 
	from p+Au reactions at 18 AGeV/c (solid symbols are the
	yield within acceptance). The lines are various parameterizations
	of the production. The right panel shows the x-distribution
	of $\pi^-$ production from p+Be at 12 GeV/c. The reactions
	with the largest number of collisions $\nu$ have the
	strongest yield at low-x.}
	\label{fig:pAparticle}

\end{figure}

Progress can be made using particle production results
from p+A experiments.
For example, the production of $\Lambda$ from p+Au
reactions increases as a function of the 
number of collisions (figure \ref{fig:pAparticle})\cite{E910pLambda}.
Folding this increase into a Glauber model,
where each projectile nucleon suffers multiple interactions, 
accounts for
approximately 70\% of the measured K$^+$ in Si+Au and Au+Au 
reactions\cite{coleqm2001}. 
The remaining increase may be driven
by complex rescattering processes that involve more than exciting
projectile nucleons.

Other results from p+A experiments can be used to 
gain more insight into the physics of rescattering. The 
measured x-distribution (x=$p_z/p_{beam}$)
of pions from p+Be is shown in figure \ref{fig:pAparticle}. Pions with
a significant fraction of the incoming momentum, $x>0.6$,
have been previously described as the 
fragmentation of the leading quark-diquark
system\cite{frag}. However the bulk of pions in pA reactions
are emitted with
$x<0.3$. These pions could be modeled as a decay of
a leading resonance, since if the decay pions are emitted 
at or near beam-rapidity, they will have a momentum fraction
$x<0.3$. 

\section{Outstanding Questions in Dense Matter}

\begin{figure}[htb]
	\centering
	\includegraphics[width=2.5in]{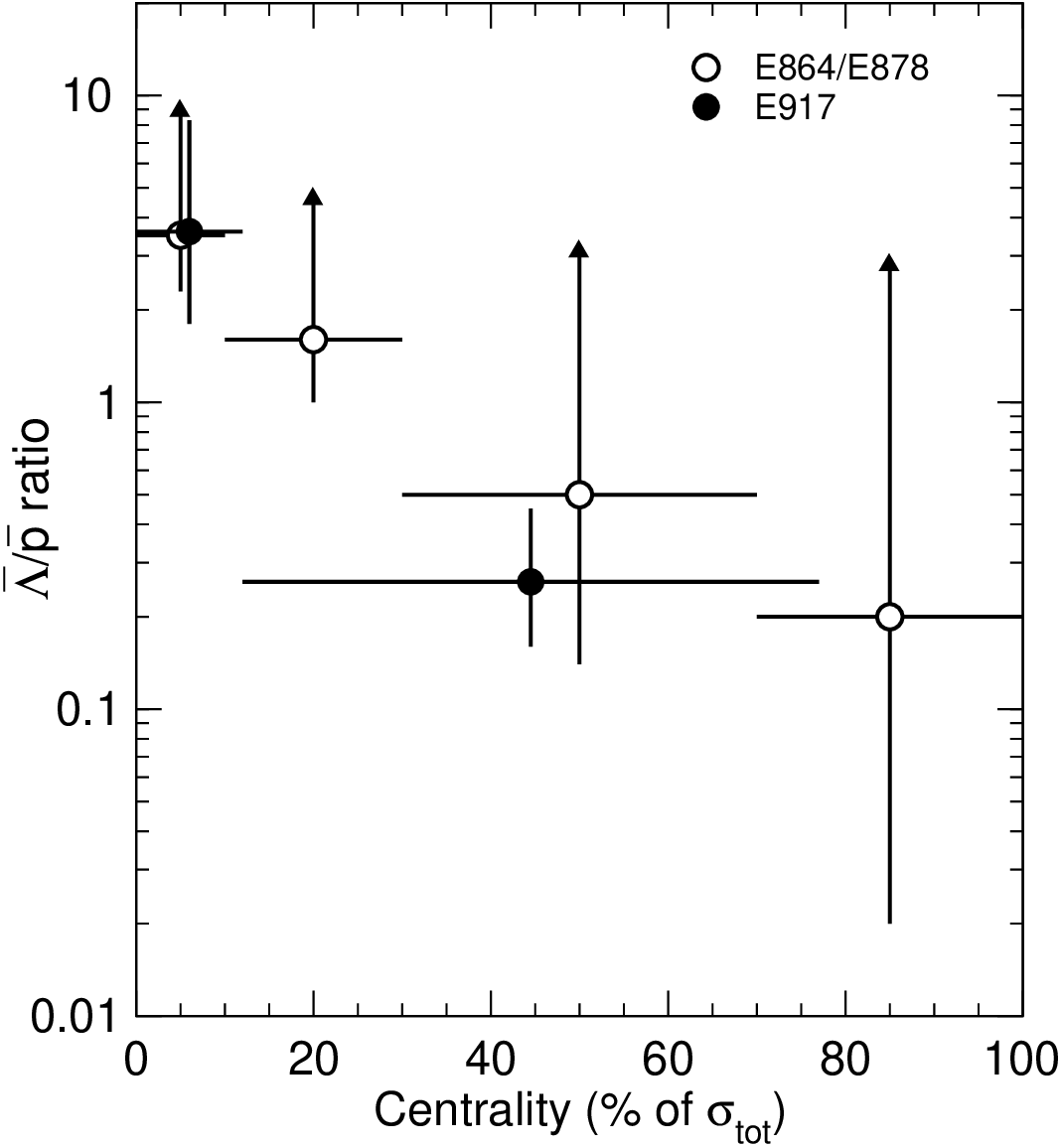}
	\caption{The measured $\overline{\Lambda}/\overline{p}$ ratio
	as a function of centrality from Au+Au at 11.7 AGeV/c
	\protect{\cite{e917lambdabar,nagle}}.}
	\label{fig:lambdabar}	
\end{figure}

One of the puzzles in the AGS program is the large 
$\overline{\Lambda}/\overline{p}$  ratio observed in both Si+Au\cite{gsfs} 
and Au+Au reactions\cite{e917lambdabar,nagle}. 
In figure \ref{fig:lambdabar} the  
$\overline{\Lambda}/\overline{p}$ ratio for central reactions rises
significantly above the ratio in peripheral reactions,
as measured by experiment E917\cite{e917lambdabar} and
inferred
from a comparison of E864 and E878 $\overline{p}$ yields\cite{nagle}.
One possible explanation is a difference in absorption
between $\overline{\Lambda}$ and $\overline{p}$. 
Models of $\overline{p}$ absorption
can now be tested against the recent anti-flow
measurements of $\overline{p}$ from E877\cite{pbarAntiFlow}.
An alternate suggestion is that n-body collisions may
increase the yield of
anti-baryons\cite{nbody}.

Looking to the future, 
there are two new planned accelerator complexes that will reach
heavy-ion beam energies of 25 AGeV. The Japan Hadron Facility
was recently funded for proton acceleration, and a
new facility is proposed for GSI, Germany.
The key physics opportunity at both will be to measure 
soft di-leptons of invariant mass
near the $\rho/\omega$ peak. This directly addresses
the question of whether hadrons change their
properties in dense matter due to
the possible change in vacuum condensate as the
nuclear density increases. 
Other opportunities include 
detailed
measurements of the centrality and beam-energy
dependence of directed and elliptic flow
to provide insight into multiple collisions and the physics
of the mean-field at high densities. It will also
be important to extend the excitation function of strangeness production
and to accurately map out $\gamma_s$ versus
E$_{beam}$.

On the more speculative side, it is an open question whether 
a method can be devised to select
events that form a very cold, yet dense system, and
explore the region
of phase space
near the conjectured color superconducting phase
of matter\cite{superconducting}. 
One possibility is to use the stochastic nature of 
heavy-ion reactions to produce event-by-event
 nuclear matter with different density and temperature. Dynamical
transport models can be used to estimate how broad this variation is
and whether there are event-by-event observables that
preferentially select reactions that pass through a cool,
dense region of phase space.
 
\section{Conclusions}

Many of the dynamical observables measured by the
AGS experiments are well
described by transport models that incorporate 
soft interactions into a 
repulsive, momentum dependent mean-field. The success list
includes directed and elliptic flow, and proton p$_t$ spectra.
In contrast, microscopic attempts to 
describe the yields of particle production
and its beam energy dependence have been relatively
unsuccessful. 
This could be resolved by first 
using the  broad range of pA data
to benchmark the detailed processes that occur
when hadrons multiply scatter, then
applying these lessons to the heavy-ion data. 

The measurements of strangeness
enhancement and extraction of strangeness chemical
saturation suggest that there is a smooth evolution of reaction
mechanism from 2-160 AGeV.
At the same time there are substantial
disagreements between data at the
AGS and current hadronic transport models.
Both facts imply that we need to improve our understanding of
the baseline properties and dynamics of dense hadronic
matter before strongly claiming the existence of a new state of
matter at the SPS.

I gratefully acknowledge conversations with G.S.F. Stephans, R. Seto,
G. Heintzelman, J.C. Dunlop, J. Lajoie, M. Rosati
K. Rajagopal, J. Cleymans, P. Danielewicz,
W. Cassing, P. Senger, B. Cole, P. Braun-Munzinger, 
C. Pinkenburg, D. Keane, S. Bass, B. Mueller,
U. Heinz, F. Becattini, A. Dumitru. This work was supported by DOE.    


\begin{thebibliography}{00}

\bibitem{qgpclaim} U. Heinz, M. Jacob, nucl-th/0002042 
%''Evidence for
%a new State of Matter: An Assessment of the Results from the CERN Lead Beam
%Programme''  
\bibitem{nbody} G. Batko, J. Randrup, T. Vetter, Nucl. Phys. {\bf A536} 786, 
(1992), Nucl. Phys. {\bf A546} 761, (1992), R. Rapp and E. Shuryak 
hep-ph/0008326, C. Greiner nucl-th/0011026
\bibitem{offmass} W. Cassing, S. Juchem, Nucl. Phys. {\bf A677} 445, 
(2000)
\bibitem{e895flow} H. Liu et al., Phys. Rev.  {\bf 84} 5488 (2000)
\bibitem{cassingMF}  W. Cassing, E.L. Bratkovskaya, S. Juchem, Nucl. Phys. 
{\bf A674} 249, (2000)
\bibitem{Sahu00}  P.K. Sahu, W. Cassing, U. Mosel, A. Ohnishi, Nucl. Phys. {\bf A672} 376, (2000)
\bibitem{e866proton} L. Ahle et al., Phys. Rev.  {\bf C 57} R466 (1998)
\bibitem{SorgeElliptic} H. Sorge, Phys. Rev. Lett. {\bf 78} 2309 (1997)
\bibitem{E895Elliptic} C. Pinkenburg et al.,   Phys. Rev. Lett. 
{\bf 83} 1295 (1999)
\bibitem{E877Elliptic} J. Barrette et al.,   Phys. Rev.  
{\bf C 55} 1420 (1997)
\bibitem{e895hbt} M. Lisa et al., Phys. Lett. {\bf B496} 1 (2000)
\bibitem{coleqm2001} B.A. Cole, these proceedings.
\bibitem{kaonAtom} E. Friedman et al.,  Phys. Rev.  {\bf C 55} 1304 (1997)
\bibitem{SchaffnerKaonMass} J. Schaffner-Bielich et al., Nucl. Phys. 
{\bf A625} 325 (1997)
\bibitem{kaonPotential} B.-A. Li and C.M. Ko,  Phys. Rev.  
{\bf C 54} 3283 (1996)
\bibitem{K0barFlow} P. Chung et al.,  Phys. Rev. Lett. {\bf85} 940 (2000)
\bibitem{e866char} L. Ahle et al.,  Phys. Rev.  {\bf C 59} 2173 (1999),
  Phys. Rev.  {\bf C 58} 3523 (1998)
\bibitem{kpiEbeam}  L. Ahle et al.,  Phys. Lett.  {\bf B476} 1 (2000),
C. Blume et al., these proceedings.
\bibitem{jcdunlop00} J.C. Dunlop and C.A. Ogilvie, Phys. Rev. {\bf C 61}
031901 (2000)
\bibitem{becattini00} F. Becattini et al., hep-ph/0011322
\bibitem{noMultipleStrange} No extensive multi-strange results exist
at the AGS, yet if the system is in relative equilibrium, then
single strange measurements provide sufficient information.
\bibitem{Odyniec}G. Odyniec, Nucl. Phys. {\bf A638} 135c (1998)
\bibitem{E910pLambda}I. Chemakin, et al., 
Phys. Rev. Lett. {\bf 85} 4868 (2000) 
\bibitem{frag} K.P. Das and R.C. Hwa,  Phys. Lett.  {\bf B68} 459 (1977)
\bibitem{gsfs} G.S.F. Stephans, J.Phys. {\bf G23} 1895 (1997)
\bibitem{e917lambdabar} B.B. Back et al., nucl-ex/0101008 
\bibitem{nagle} T.A. Armstrong et al., , Phys. Rev. {\bf C 59}
2699 (1999)
\bibitem{pbarAntiFlow} J. Barrette et al., Phys. Lett.  {\bf B485} 319 (2000)
\bibitem{superconducting} M. Alford, K. Rajagopal, F. Wilczek,
Phys. Lett.{ \bf B422} 247 (1998), 
J. Berges, K. Rajagopal, Nucl. Phys. {\bf B538} 215,(1999)
\end{thebibliography}
\end{document}